\documentstyle[11pt,newpasp,twoside,psfig]{article}
\begin{document}

\title{Polarized CMB: Reionization and Primordial Tensor modes}

\author{Asantha Cooray}

\affil{Theoretical Astrophysics, California Institute of Technology, Pasadena, California 91125. E-mail: asante@caltech.edu}



\begin{abstract}
We discuss upcoming opportunities with cosmic microwave background (CMB) observations during the post-WMAP era.
The curl-modes of CMB polarization probe inflationary gravitational waves (IGWs).
While a significant source of confusion is expected from cosmic shear conversion of polarization
related to density perturbations, higher resolution observations of CMB anisotropies can be used for a lensing
reconstruction and to separate gravitational-wave polarization signature from that of lensing. 
Separations based on current lensing reconstruction techniques allow the possibility to probe inflationary energy scales below
 10$^{15}$ GeV in a range that includes grand unified theories. 
The observational detection of primordial curl-modes is aided by rescattering at late times during the reionized epoch
with optical depth to electron scattering at the level of 0.1 and above. 
An improved measurement of this optical depth is useful to optimize experimental parameters of a post-WMAP mission attempting
to target the IGW background.
\end{abstract}


\section{CMB: At Present}
The cosmic microwave background (CMB) is now a well known probe of the early universe. 
The temperature fluctuations in the CMB, especially the so-called acoustic peaks in the angular power spectrum of CMB anisotropies, 
capture the physics of primordial photon-baryon fluid undergoing oscillations in the potential 
wells of dark matter (Hu et al. 1997). The associated physics --- involving the evolution of a single
photon-baryon fluid under Compton scattering and gravity --- are both simple and linear, and
many aspects of it have been discussed in the literature since the early 1970s (Peebles \& Yu 1970; Sunyaev \& Zel'dovich 1970).
The gravitational redshift contribution at large angular scales (e.g., Sachs \& Wolfe 1968) and the photon-diffusion damping
at small angular scales (e.g., Silk 1968) complete this description.

By now, the structure of the first few acoustic peaks 
is well studied with NASA's Wilkinson Microwave Anisotropy Probe (WMAP) mission (e.g., Spergel et al. 2003), while 
 in the long term, ESA's Planck surveyor\footnote{http://astro.estec.esa.nl/Planck/}, 
will extend this to a multipole of $\sim$ 2000 with better frequency coverage.
Beyond temperature fluctuations, detections of the polarization anisotropy, related to density or scalar
perturbations, have now been made by the DASI experiment (Kovac et al. 2003), followed up with  
an improved study of the temperature-polarization power spectrum with WMAP (Kogut et al. 2003).

An interesting result from WMAP data is related to the very large angular scale polarization signal, which probes
the local universe and associated astrophysics instead of physics at the last scattering surface probed with polarization at
tens of arcminute scales. This local universe contribution arises when the universe reionizes again and
the temperature quadrupole begins to rescatter to produce a new contribution
to the polarization anisotropy at angular scales corresponding to the horizon at the new scattering surface 
(Zaldarriaga 1997). Such a signal was measured in the WMAP temperature-polarization cross-correlation
 power spectrum and was interpreted as rescattering with an  optical depth to electron scattering, inferred from the
amplitude of the large angular scale CMB polarization, of $0.17 \pm 0.04$ (Kogut et al. 2003).

Such an optical depth suggests early reionization; for example, if the universe reionized completely and instantaneously at
some redshift, the measured optical depth suggests a reionization redshift of $\sim$ 17 ($\pm 5$). 
Such a high optical depth, in the presence of
a $>$ 1\% neutral Hydrogen fraction from z $\sim$ 6 Sloan quasars (Fan et al. 2002),
 suggests a complex reionization history. A complex, and patchy, 
reionization, however, is expected if the reionization is dominated by the
UV light from first luminous objects, though to explain the high level of the WMAP's optical depth requires
high star formation efficiency at redshifts of order 25 with a possibility for a population of metal-free massive Pop III
stars (e.g., Cen 2003).

In general, one can only extract limited information related to the reionization history 
from  polarization peak at tens of degrees or more angular scales (e.g., Hu \& Holder 2003).
One reason for this is the large cosmic variance associated with 
measurements at multipoles of $\sim$ 10. To extract detailed information of the reionization process,
one can move to temperature fluctuations generated during the reionization era at angular scales
corresponding to few arcminutes and below. At these small angular scales, potentially interesting contributions from
high redshifts include a scattering contribution related to moving electrons in ionized patches and
fluctuations in the moving electron population (e.g., Santos et al. 2003),
and a thermal Sunyaev-Zel'dovich effect, just as  in massive galaxy clusters containing hot electrons (Sunyaev \& Zel'dovich 1980),
related to the first supernovae bubbles (Oh et al. 2003). The same redshift ranges are expected to be
illuminated in the near-infrared band between 1 to 3 $\mu$m in the form of spatial fluctuations to the cosmic
IR background (Cooray et al. 2003). Potentially interesting studies to extract more information related to the
reionization process include cross-correlation studies between these small scale
effects in CMB as well as other wavelengths. Upcoming experiments such as the South Pole Telescope and the
Atacama Cosmology Telescope will provide some of the first possibilties in this direction.

\begin{figure}[!h]
\centerline{\psfig{file=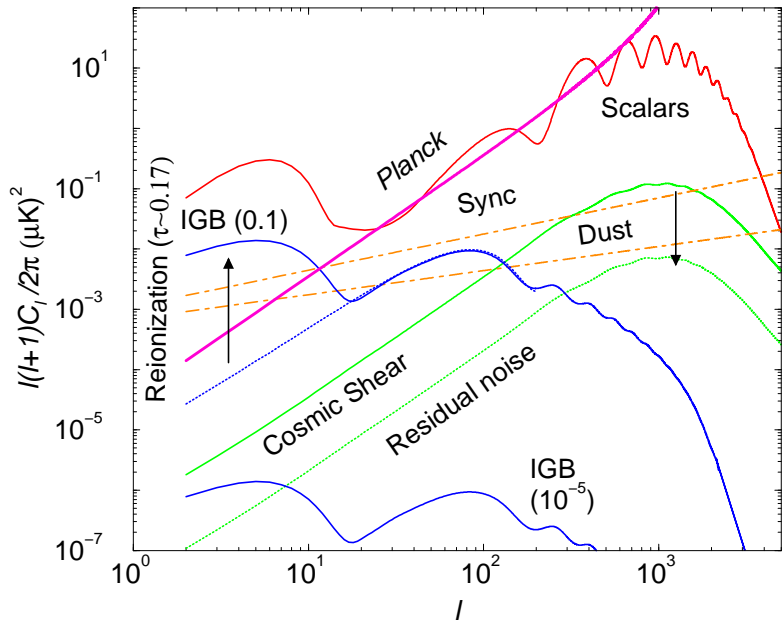,height=20pc,angle=-90}}
\caption{CMB polarization anisotropies in the gradient (E) mode due to scalars and curl (B) modes due to
the inflationary gravitational wave background (IGB, with the number in parenthesis showing the normalization
in terms of the tensor-to-scalar ratio, which in terms of the energy scale of inflation is
$2 \times 10^{16}$ GeV (0.1), top curve, and $2 \times 10^{15}$ GeV (10$^{-5}$), bottom curve). 
Note the effect due to
reionization where rescattering produces new large angular scale anisotropies. 
For optical depths at the level of 0.1, the peak of the power related to IGB curl modes is at tens of degree angular scales and not
at the degree scale ($l \sim 100$) corresponding to the bump at recombination.
Note that this enhancement, amounting to over two magnitudes in power at $l \sim$ few 
is significant and will dominate the potential detectability of the IGB via polarization.
The IGB detection is confused with cosmic shear conversion of  a fractional E-mode $\rightarrow$ B-mode 
by the intervening large scale structure. 
The residual noise curve related to cosmic shear is after doing lens-cleaning in polarization maps
with an arcminute scale CMB experiment with instrumental polarization sensitivity 
at the level of 1 $\mu$K $\sqrt{\rm sec}$. For reference, we show an estimated level of contamination from
dust and synchrotron emission (at 150 GHz);  this noise level can be improved with foreground cleaning
in multifrequency data with sensitivity better than the foreground noise contributions. 
The curve labeled Planck is the noise curve related to upcoming Planck (HFI) polarization observations as
a function of each multipole.
}
\label{fig:cl}
\end{figure}

\section{CMB: As a Probe of the Inflation}

\begin{figure}[!ht]
\centerline{
\psfig{file=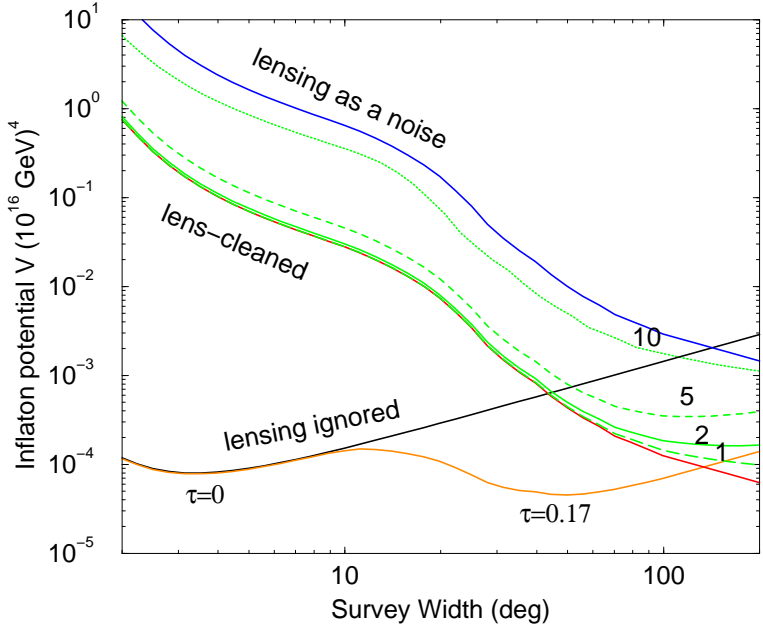,height=20pc,angle=-90}}
\caption{Minimum inflation potential observable at
     $1\sigma$ as a function of survey width. The panel shows an experiment
     with $1\, \mu{\rm      K}~\sqrt{\rm sec}$ sensitivity.
     The bottom curve shows results assuming no lensing confusion
for the two cases with $\tau=0$ and $\tau=0.17$; for $\tau=0$, power is at degree scales and
can be extracted from a survey that targets a few square degree area. In the case of $\tau=0.17$, 
one improves the limit on IGB amplitude slightly by targeting a larger area. 
The top curves show results including the effects
     of an unsubtracted lensing or with a partly subtracted lensing based on lens-cleaning methods using
     quadratic statistics; we take $\theta_{\rm     FWHM}=10,5,2$ and 1 arcminute from top to bottom, while the
     curve below 1 arcminute assumes  lensing subtraction  with no instrumental noise.
As shown in Fig.~1, the recombination peak at
degree scales is confused substantially by the lensing confusion but the reionization peak at tens or more degree scales is
less affected. Thus, a poor resolution experiment with no capability for lens-cleaning can be optimized
with a better understanding of the optical depth to reionization, if the achievable detector noise sensitivities 
are such that lens cleaning at higher resolution is unlikely to make a substantial improvement.}
\label{fig:bmodes}
\end{figure}

\begin{figure}[t]
\centerline{
\psfig{file=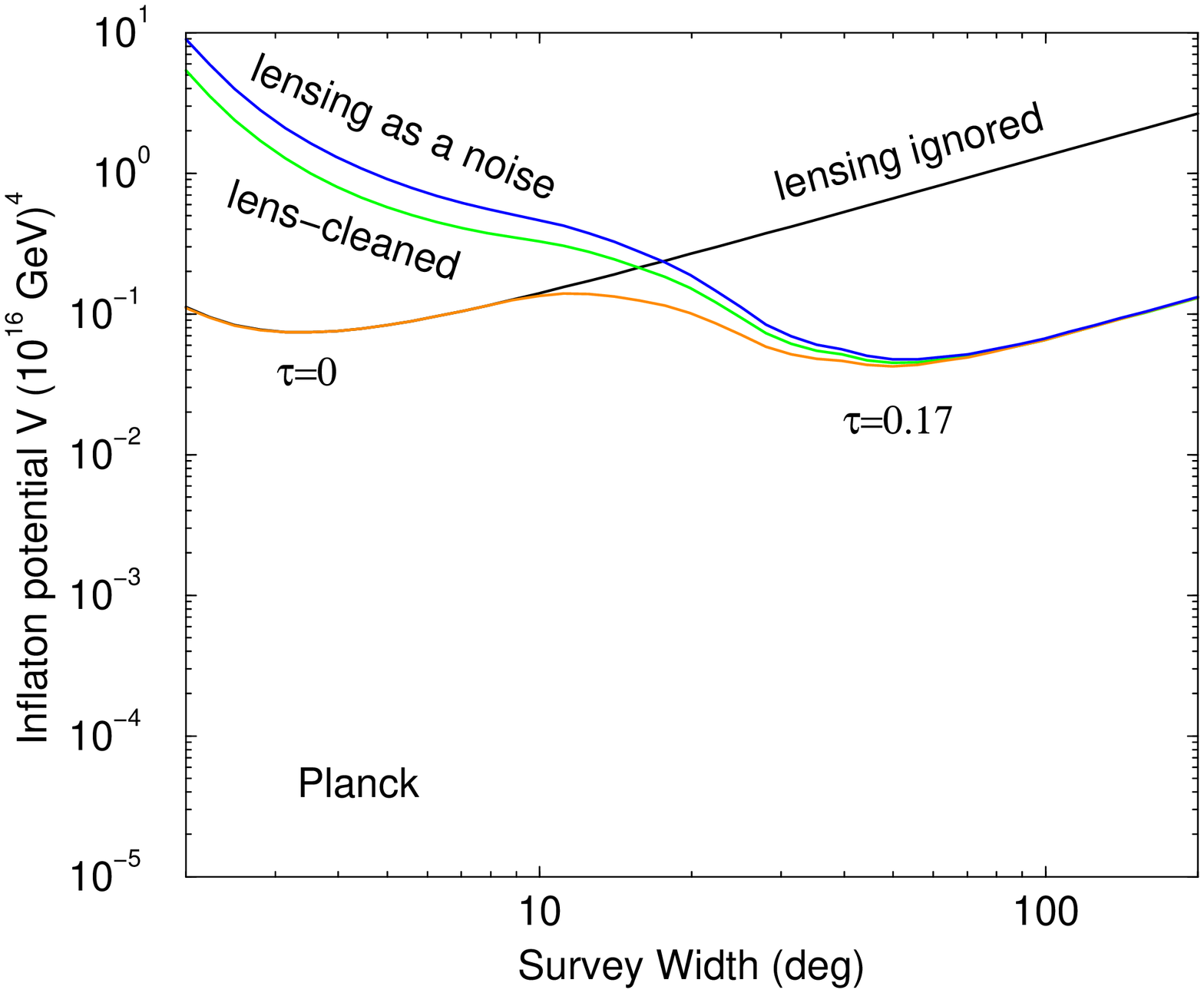,height=20pc,angle=-90}}
\caption{Same as figure 2, but with sensitivity for polarization observations at the level expected from Planck (HFI) with
the same resolution as Planck polarization maps. The importance of reionization is now clearer: given the poor
sensitivity and resolution, lens-cleaning is not important here and with $\tau=0.17$ one achieves the same limit
by targeting a small area than the whole sky, but if $\tau$ were to be smaller, the achievable limit is higher due to
lensing confusion of the degree scale recombination peak.}
\label{fig:bmodesplanck}
\end{figure}

Though acoustic oscillations in the temperature
anisotropies of CMB suggest an inflationary origin for
primordial perturbations, it has been argued for a while
 that the smoking-gun signature for inflation would be the detection of a stochastic
background of gravitational waves (e.g., Kamionkowski \& Kosowsky 1999).  These gravitational-waves produce a distinct signature in the
CMB in the form of a contribution to the curl, or magnetic-like, component of the polarization (Kamionkowski et al. 1997; Seljak \& Zaldarriaga 1997).
Note that there is no contribution from the dominant scalar, or density-perturbation, contribution to these curl modes.

In figure~1, we show the contribution from dominant scalar modes to the polarization in the gradient
 component and from gravitational-waves to the curl polarization. Note that the amplitude of the IGB contribution is
highly unknown and simply depends on the energy scale of the inflation. The current limit from CMB anisotropy data
suggests an upper limit of order $2 \times 10^{16}$ GeV (Melchiorri \& Odman 2003).

An important issue related to potential detectability of the curl-modes related to gravity waves is the
peak of its contribution; if the universe did not reionize, the peak of the contribution is at the multipoles of $\sim$ 100
corresponding to the contribution generated at the recombination.
On the other hand, in the presence of a reionization, the rescattering adds extra power at large angular scales. For optical
depths of order 0.17, as estimated by WMAP, the peak is now at tens or more degree scales. This substantial increase
in the amplitude aids potential detection of the curl modes and to a large extent will determine the
observational details of a post-WMAP mission.  For example, if a substantial reionization exists, one can target the
large angular scale polarization with a poor resolution experiment.
If instrumental noise sensitivities can be substantially improved it may be better to target both reionized 
contribution at large angular scales and the recombination contribution at small angular scales to substantially improve
the limiting amplitude of the IGB. For this, a major source of concern is the confusion from effects that generate curl-modes
and not associated with the IGB. 

Among these, the few percent conversion of E to B from cosmic shear modification to the
polarization pattern by the intervening large scale structure is important (Zaldarriaga \& Seljak 1998). These
lensing-induced curl modes introduce a noise from which gravitational waves must be distinguished in the CMB polarization.
We now discuss this contribution and how it can be reduced by performing weak lensing studies in high resolution CMB 
temperature and polarization maps.

{\bf Weak Gravitational Lensing with CMB:} 
Gravitational lensing of the CMB photons by the mass fluctuations in the 
large-scale structure is now well understood (Hu 2001). 
Since the lensing effect involves the angular gradient of CMB photons and leaves the surface brightness unaffected, its  
signatures are at 
the second order in temperature. Effectively, lensing smooths the acoustic peak structure at large angular scales and
moves photons to small scales. When the CMB gradient at small angular scales are lensed by foreground structures 
such as galaxy clusters, new anisotropies are generated at arcminute scales. For favorable cosmologies, the mean gradient is of order
$\sim$ 15 $\mu$K arcmin$^{-1}$ and with deflection angles or order 0.5 arcmin or so from massive clusters, the lensing effect results
in temperature fluctuations of order $\sim$ 5 to 10 $\mu$K. 

One can effectively extract this lensing contribution, and the
integrated dark matter density field responsible for the lensing effect, via 
 quadratic statistics in the temperature and the polarization (Hu 2001; Hu \& Okamoto 2003; Cooray \& Kesden 2003; Kesden et al 2003) 
or likelihood-based techniques that are optimized for lensing extraction (Hirata \& Seljak 2003). 
We show errors for a construction of convergence from CMB temperature, with arcminute-scale angular resolution,
in figure~4 and a comparison to what can be achieved with  weak lensing studies involving galaxy shapes. 
Note that one probes linear clustering with CMB while most sensitive measurements at low redshifts,
with galaxy shear measurements, are in the non-linear regime. This makes a big difference in terms of cosmological
parameter extraction since the uncertainty related to understanding non-linear clustering properties,
in the presence of a dark energy or massive neutrino, will limit the extent to which one can extract information
from low redshift lensing surveys. On the other hand, we suggest the importance of combining the two: CMB, with
a source at z $\sim$ 1100, and galaxy surveys, with sources at $z \sim 2$, probe complimentary information
and the evolution of dark matter clustering between the two redshift ranges.

When considering the effect of a lensing confusion to IGB detection, it is desirable to make use of CMB data for a lensing
reconstruction. This is due to the fact that what one extract from galaxy shape measurements out to a redshift of a few,
at most, only leads to a partial accounting of the total effect
(the difference in curves in figure~4).  By mapping lensing deflection from CMB data, 
the lensing effect on polarization can be corrected to reconstruct the intrinsic CMB
polarization at the surface of last scattering. Now the only curl component would be that due to gravity waves.
In figure~1, we show how well the lensing signal can be extracted. With no noise all-sky maps, and using quadratic statistics,
one can remove the lensing
contribution with a noise contribution that is roughly an order of magnitude lower than the original confusion.
The extent to which the IGB amplitude limit, in terms of the energy scale of inflation,
 can be improved in the presence of cosmic shear confusion is summarized in Fig.~2 for  a 1 $\mu$K $\sqrt{\rm sec}$ sensitivity
experiment, while in Fig.~3, we consider the same for a 
$\sim$  25 $\mu$K $\sqrt{\rm sec}$ sensitivity experiment with resolution similar to Planck. 
Following Kesden et al. (2002; Knox \& Song 2002), if there is no instrumental-noise limitation, the sensitivity to gravity-wave
signal is maximized by covering as much sky as possible and allow the accessibility to an inflaton potential of $\sim 2 \times 
10^{15}$ GeV.
This  limit is slightly improved with likelihood techniques which extract additional information beyond quadratic statistics 
(Seljak \& Hirata 2003). If the optical depth is to be high at the level of 0.1, then limits at the same level or
slightly worse can be achieved by targeting the degree scale peak with a poor resolution experiment with
no need for a removal of the confusion related to cosmic shear.
Thus, in conclusion, we iterate the importance of reionization 
and we suggest that an improved measurement of the optical depth will be helpful to optimize a
post-WMAP mission related to the IGB detection as part of NASA's Einstein Probes program.

\begin{figure}[t]
\centerline{
\psfig{file=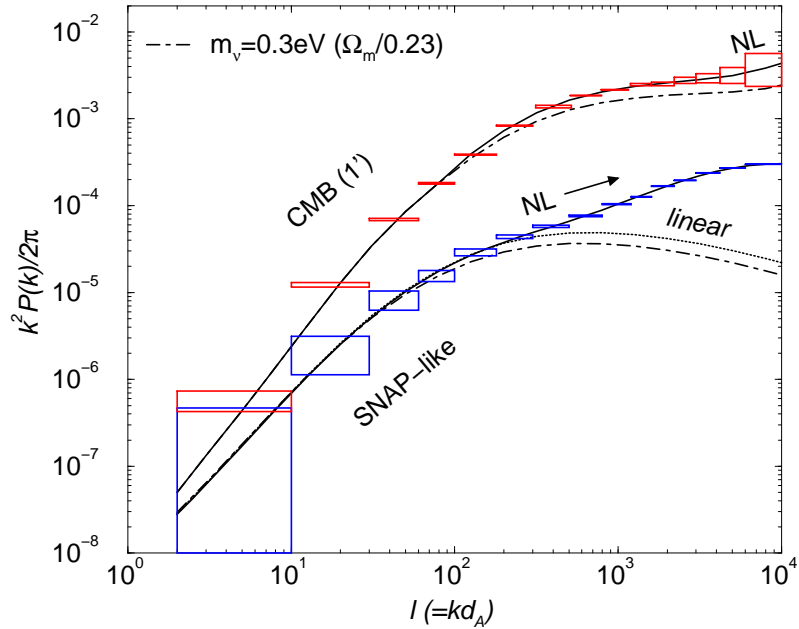,height=20pc,angle=-90}}
\caption{The lensing convergence reconstruction with CMB, with arcminute-scale angular resolution,
 and galaxy shape surveys, with sensitivity similar to a space-based campaign such as the one planned with SNAP.
Note that the convergence measurements at low redshifts are mostly in the non-linear regime of dark matter clustering
while with CMB, one mostly reconstructs linear clustering of matter. While there are other advantages such as the precise
location of the source ($z \sim 1100$) with CMB related lensing studies, 
the two lensing surveys with CMB and galaxies provide complimentary information and can be combined to improve
the determination of certain cosmological parameters. We consider one case here with a massive neutrino, which can
be distinguished at many tens of sigmas if the mass is at the level of 0.1 eV.}
\label{fig:cl}
\end{figure}

{\it Acknowledgments:}
The author thanks Jamie Bock, Brian Keating, Marc Kamionkowski, Mike Kesden and Peng Oh for collaborative work, 
and acknowledges support from the Sherman Fairchild foundation and the Department of Energy. The author thanks
National Science Foundation for travel support to attend the IAU General Assembly.


\end{document}